\pdfoutput=1
\documentclass{aa}

\usepackage{graphicx}
\usepackage{txfonts}

\newcommand\ha{H$\alpha$}
\newcommand\ca{\mbox{Ca\,{\sc ii}}}
\newcommand\ti{\mbox{Ti\,{\sc ii}}}
\newcommand\si{\mbox{Si\,{\sc i}}}
\newcommand\fe{\mbox{Fe\,{\sc i}}}
\newcommand{\kms}{km\,s$^{-1}$}
\newcommand{\flux}{$\rm erg\,s^{-1}\,cm^{-2}$}

\begin{document}

\title{Impulsive wave excitation by rapidly changing granules}

\author{Hannah Kwak\inst{1}
  \and Jongchul Chae\inst{1}
  \and Maria S. Madjarska\inst{1}
  \and Kyuhyoun Cho\inst{1}
  \and Donguk Song\inst{2}}

\offprints{jcchae@snu.ac.kr}
\institute{Astronomy Program, Department of Physics and Astronomy, Seoul National University, Seoul 08826, Republic of Korea
  \and
  National Astronomical Observatory of Japan, 2-21-1 Osawa, Mitaka, Tokyo 181-8588, Japan}

\date{Received date, accepted date}

\abstract
{It is not yet fully understood how magnetohydrodynamic waves in the interior and atmosphere of the Sun are excited. Traditionally, turbulent convection in the interior is considered to be the source of wave excitation in the quiet Sun. Over the last few decades, acoustic events observed in the intergranular lanes in the photosphere have emerged as a strong candidate for a wave excitation source. Here we report our observations of wave excitation by a new type of event: rapidly changing granules. Our observations were carried out with the Fast Imaging Solar Spectrograph in the \ha\ and \ca\,8542\,\AA\ lines and the TiO 7057\,\AA\ broadband filter imager of the 1.6\,m Goode Solar Telescope at the Big Bear Solar Observatory. We identify granules in the internetwork region that undergo rapid dynamic changes such as collapse (event~1), fragmentation (event~2), or submergence (event~3). In the photospheric images, these granules become significantly darker than neighboring granules. Following the granules' rapid changes, transient oscillations are detected in the photospheric and chromospheric layers. In the case of event~1, the dominant period of the oscillations is close to 4.2\,min in the photosphere and 3.8\,min in the chromosphere. Moreover, in the \ca--0.5\,\AA\ raster image, we observe repetitive brightenings in the location of the rapidly changing granules that are considered the manifestation of shock waves. Based on our results, we suggest that dynamic changes of granules can generate upward-propagating acoustic waves in the quiet Sun that ultimately develop into shocks.}

\authorrunning{Kwak et al.}
\titlerunning{Impulsive Wave Excitation by Rapidly Changing Granules}

\maketitle

\section{Introduction} \label{intro}

A wide range of observations have revealed that oscillations and waves are abundant in the solar atmosphere. They are clearly observed in the umbral and penumbral regions of sunspots \citep[e.g.,][]{bec72} and the network and internetwork regions of the quiet Sun \citep[e.g.,][]{orr66, deub90}. It is not yet known how waves are generated in the solar atmosphere. Theoretical studies suggest that solar acoustic waves can be produced by impulsive disturbances in a gravitationally stratified medium \citep{kal94, chae15}. \citet{chae15} report that when a region is disturbed by an impulsive event, acoustic waves with an acoustic cutoff frequency naturally arise in a medium. In terms of global p-mode oscillations, it is now generally accepted that p-modes are excited by turbulent convection \citep{gol90}. \citet{nig99} found that a wave excitation source is located at a depth of 75$\pm$25\,km below the photosphere by comparing theoretical and observed p-mode power spectra. Since turbulent convection occurs ubiquitously, the observed oscillations show the superposition of oscillation signals coming from different sources. In this regard, investigating an individual wave excitation event that is well-separated in time and space could facilitate the establishment of the wave excitation process.

The wave excitation process appears to be related to localized disturbances below the photosphere. Using one-dimensional simulations, \citet{goo92} showed that acoustic waves are excited by individual wave excitation events occurring less than 200\,km below the base of the photosphere. \citet{rim95} found a spatially localized transient wave energy flux that arises due to the excitation of waves beneath the photosphere. These phenomena are termed acoustic events and they are generally found in the intergranular lanes \citep{rim95, bel10}. On a much larger scale, they are found in intergranular lanes located in or near the boundaries of regions with predominant downward vertical motions and horizontal converging flows on a mesogranular scale \citep{mal15}. Before the acoustic events occur, darkening of intergranular lanes is observed at the photospheric level, and the darkening is interpreted as catastrophic cooling that occurs in the intergranular lanes below the photosphere \citep{rim95}. Similarly, theoretical studies suggest that localized cooling events and subsequent downflows could be related to a wave excitation process \citep{rast99, ska00}.

Furthermore, the waves generated beneath the photosphere propagate upward and affect the upper atmosphere. In the internetwork region, small intermittent brightenings have been observed in \ca~H and K filtergrams and time series of spectra \citep[e.g.,][]{bap71, zir74, cram83}. These brightenings are called \ca\ bright grains or internetwork grains, and several studies have suggested that they are closely related to wave phenomena due to their recurrent behavior \citep[for a review, see][]{rut91}. Based on one-dimensional non-local thermodynamic equilibrium (non-LTE) radiation-hydrodynamic simulations, \citet{car97} asserted that the bright grains are produced by waves coming from below the photosphere that subsequently develop into acoustic shocks in the mid-chromosphere ($\sim$1\,Mm above $\tau_{500}= 1$). Several observational studies have shown that most of the chromospheric bright grains coincide with photospheric oscillations and even photospheric darkening events. \citet{hoe02} reported that sites with the largest acoustic flux (i.e., acoustic events) in the photosphere tend to colocate with chromospheric bright grains with an average time delay of 2\,min. Similarly, \citet{cad03} investigated 1527 G-band darkening events and found that 72\,\% of the photospheric darkenings are followed by \ca~K brightenings with time lags of about 2\,min. Using time series of \ca~H, \ha, and \fe\ spectra, \citet{kam06} established that the \ca~H bright grains are caused by acoustic shocks in the chromosphere that are associated with enhanced 5~mHz oscillations in the photosphere.

Recently, space-based observations that do not suffer from the atmospheric seeing and cover a large field of view (FOV) permitted the simultaneous identification of numerous acoustic events. Using high-resolution data taken with the Solar Optical Telescope (SOT) on board the Hinode satellite, \citet{mal12} studied the properties of acoustic events such as acoustic flux, velocity amplitude, continuum intensity, and divergence. They also found that most acoustic events appear in the intergranular lanes in regions with downward velocities and converging motions. Even though general properties of acoustic events are well known from the analysis of large FOV high-resolution data, the wave excitation process of individual acoustic events has not yet been investigated in detail.

In this paper, we report the excitation of waves by a new type of event: granules that undergo rapid dynamic changes. These events differ from classic acoustic events. The new acoustic events studied here are caused by granules that experience dynamic changes rather than by events occurring in intergranular lanes. These granules either collapse, fragment, or submerge in the photospheric layer. To understand the wave excitation and propagation, we investigated the temporal behavior of the granules as well as the Doppler velocities in the photospheric and chromospheric layers. We used high-spatial and high-spectral resolution data obtained by the Fast Imaging Solar Spectrograph \citep[FISS;][]{chae13} and high-resolution photospheric data taken with the TiO 7057\,\AA\ broadband filter imager \citep{cao10} installed on the 1.6\,m Goode Solar Telescope (GST) of the Big Bear Solar Observatory. Our analysis shows a one-to-one correlation between dynamically changing granules and the excitation of acoustic waves.

\section{Observations and data analysis} \label{obs}

The observations were taken on June 14, 2017, in a quiet Sun region with the FISS and the TiO 7057\,\AA\ broadband filter imager. The FISS is a dual-band Echelle spectrograph that takes simultaneous \ha\ and \ca\ 8542\,\AA\ spectrograms with imaging capability. The spatial pixel size is 0\farcs16, and the spectral pixel size is 0.019\,\AA\ in the \ha\ band and 0.026\,\AA\ in the \ca\ band. The FOV, with a size of 24\arcsec\ $\times$ 40\arcsec, is centered at the solar heliographic coordinates X = $-140$\arcsec\ and Y = $-70$\arcsec. It  was scanned with a cadence of 27\,s for one hour, from 17:52:03 to 18:52:13\,UT (hereafter the time 17:52:03\,UT is referred to as $t=0$). Detailed information about the instrument and the basic data processing, including the flat-field correction, distortion correction, and noise suppression, is given in \citet{chae13}.

\begin{figure}[ht!]
\centering
\includegraphics[width=\hsize]{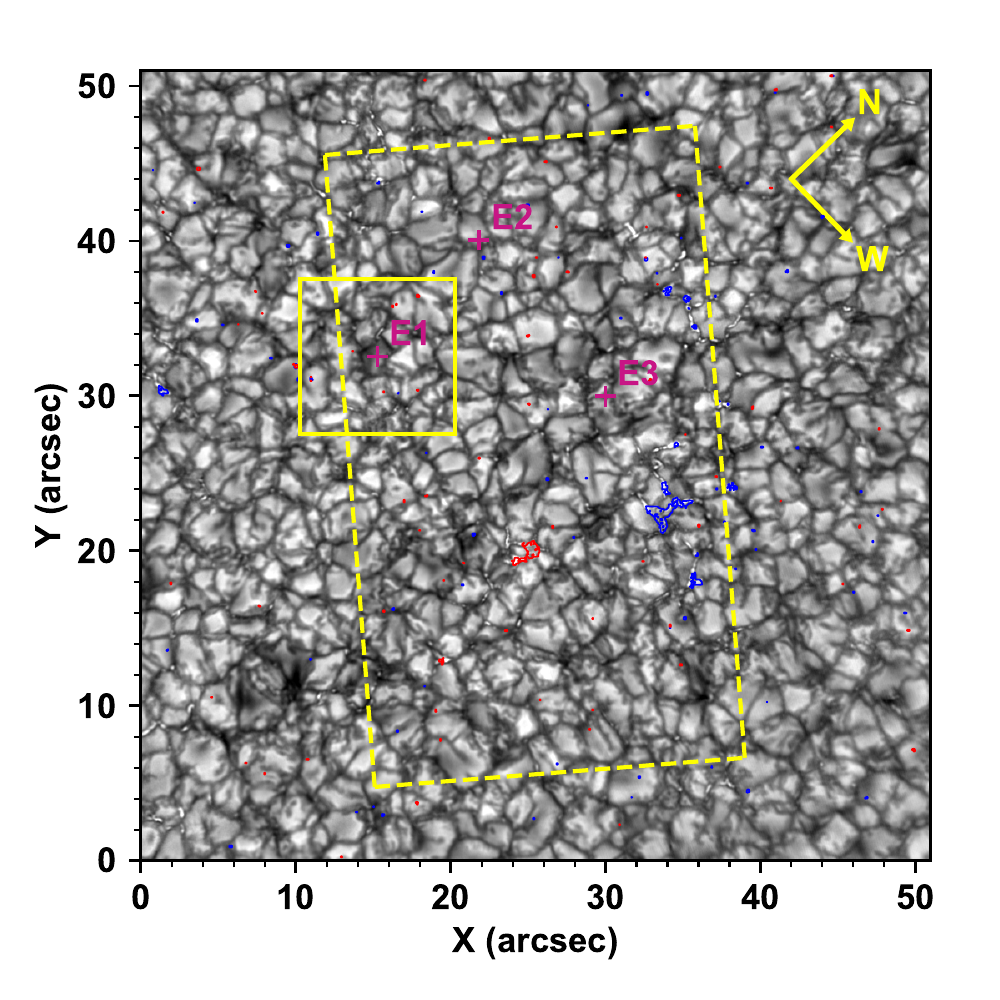}
\caption{Photospheric TiO 7057\,\AA\ broadband filter image of a quiet Sun region taken at 18:03:56\,UT ($t=11.9$\,min). The dashed yellow line rectangle represents the FOV of the FISS. The small yellow square indicates one of our regions of interest, and the size of the square is 10\arcsec$\times$10\arcsec. The locations of each event are marked with purple cross symbols and annotated as E1, E2, and E3. The contours represent the line of sight (LOS) magnetogram obtained from the Near-Infrared Imaging Spectro-polarimeter (NIRIS) installed on the GST. The blue and red contours represent LOS magnetic field strengths of $-100$\,G and $100$\,G, respectively. \label{fig1}}
\end{figure}

We acquired photospheric images using the TiO 7057\,\AA\ broadband filter imager at 15\,s cadence and an exposure time of 0.7\,ms. The speckle reconstruction was done with the Kiepenheuer-Institut Speckle Interferometry Package \citep{wog08} to achieve diffraction-limited images. The spatial pixel size is 0\farcs034 and the FOV is 70\arcsec\ $\times$ 70\arcsec. Figure~\ref{fig1} shows the TiO image taken at 18:03:56\,UT ($t=11.9$\,min). The observed region includes an internetwork region and some magnetic elements that are considered to be part of the network region. The TiO and the FISS data were aligned using the TiO and the FISS \ha--4\,\AA\ raster images.

We obtained the line-of-sight Doppler velocities using the lambdameter method \citep{deub96}. We set $\Delta\lambda$ and determined the $\lambda_m$ that satisfies the equation $I(\lambda_m-\Delta\lambda) = I(\lambda_m+\Delta\lambda)$. Since the \ha\ and \ca\ 8542\,\AA\ bands also include photospheric lines such as \ti\ 6560\,\AA\ and \si\ 8536\,\AA, we could also measure the photospheric velocities. In this study, the \ti\ 6560\,\AA\ and \ha\ lines were used to infer the photospheric and chromospheric velocities, respectively. In addition, we only obtained velocity signals with periods from 1 to 6\,min by applying a bandpass filtering method based on the wavelet analysis given by \citet{tor98}, which eliminates the solar granulation pattern corresponding to an approximately 10\,min period and noise. The bandpass filter we applied is found to be broad enough not to lose any of the acoustic events' power.

In a wavelet analysis, it is important to use an appropriate background noise model for detecting significant oscillatory power. To evaluate the background noise model from the FISS data, we took a 16\arcsec\ $\times$ 32\arcsec\ area of the central region from the \ha\ and \ti\ data and obtained spatially averaged Fourier power spectra, which are regarded as the background noise model $\sigma(\nu)$. The Fourier power spectra were modeled using the following combination of a power law function and a Gaussian function:

\begin{equation}
\sigma(\nu) = A\nu^s+B \exp \left(-\frac{(\nu-\nu_0)^2}{\sigma_G^2}\right).
\end{equation}

The model has five parameters for each line. The inferred parameters are $A= 0.02, \, s = -0.79, \,B = 7.1, \, \nu_0 = 4.3 \times 10^{-3}\,\rm{Hz}, $ and$  \, \sigma_G = 3.9 \times 10^{-3}\,\rm{Hz}$ for \ha, and $A= 2.1 \times 10^{-5},  \, s = -1.5,  \, B = 0.66,  \, \nu_0 = 3.6 \times 10^{-3}\,\rm{Hz}, \,$and $\sigma_G = 9.0 \times 10^{-4}\,\rm{Hz}$ for \ti. The results of the noise model fitting are given in Appendix~\ref{model}. We then calculated the 95\,\% local confidence level following \citet{auc16} and \citet{kay20}. We note that the noise model considered here is not for measurement errors, but for the background signals commonly seen in the data. The Gaussian component in this noise model represents the 5~min p-mode oscillations. Therefore, the events above a significance level are those that have much stronger power than the p-modes.

\begin{figure*}[ht!]
\centering
\includegraphics[width=\hsize]{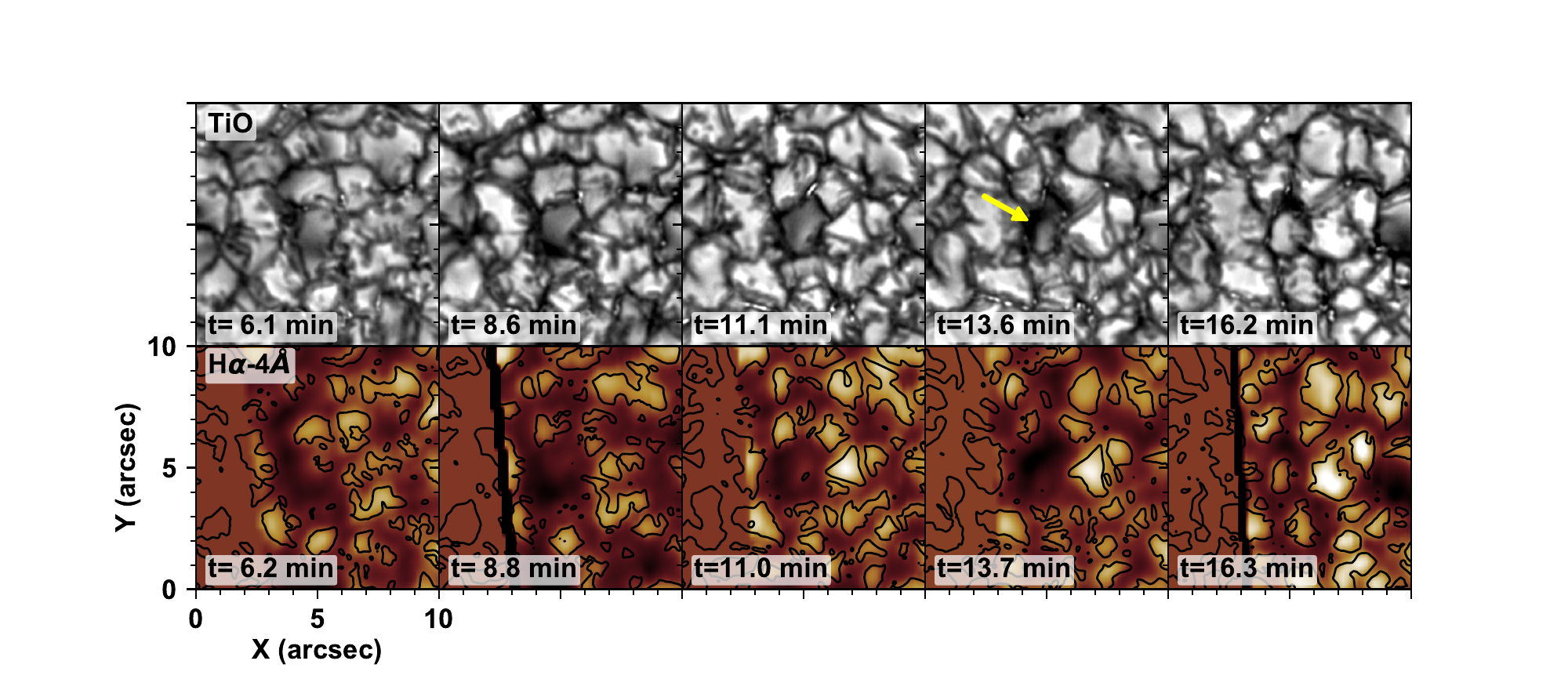}
\caption{Temporal evolutions of the granule of event~1 in the TiO images (upper panel) and the \ha--4\,\AA\ raster images (lower panel). The contours on the \ha--4\,\AA\ raster images represent the granules obtained from the TiO images. The yellow arrow indicates the position where the dynamic changes of the granule occurred. \label{th1}}
\end{figure*}

The wave energy flux is calculated within the 3\,min band (2 to 4\,min) at each spatial pixel by following the method and assumptions given in \citet{chae17}. We assumed that the acoustic waves propagate vertically in a gravitationally stratified medium (from the photosphere to chromosphere) and that the medium is isothermal. Since waves, in reality, may propagate in the non-vertical directions as well, only some part of the total energy may be carried vertically, and  the energy flux in the chromosphere may be smaller than that in the photosphere. Therefore, we estimated the wave energy flux in the photospheric level. The photospheric wave energy flux is given by the expression
\begin{equation}
F_{w}=U_{1}v_{g}=\rho_{1}P_{\nu,1}\frac{c_s^2}{v_{p}}
=\rho_{1}P_{\nu,1}\frac{c_{s,1}c_{s,2}}{v_{p}},
\end{equation}
where $U$ is the wave energy density, $v_{g}$ is the group velocity, $\rho$ is the mass density, $P_{\nu}$ is the 3\,min band oscillation power, ${c_s}$ is the sound speed, and $v_{p}$ is the phase velocity. The photospheric and chromospheric parameters are denoted by 1 and 2, respectively. We determined the group velocity $v_{g}$ from the phase velocity $v_{p}$ using the relationship of $v_{g}v_{p} = c_s^2$, which is applicable to acoustic waves. The phase velocity $v_{p}$ can be expressed as
\begin{equation}
v_{p}=\frac{\omega}k=\omega\frac{\Delta z}{\Delta\Phi},
\end{equation}
where $\omega$ is the frequency, $k$ is the wavenumber, $\Delta\Phi$ is the phase difference, and $\Delta z$ is the height difference. The value of $\Delta\Phi$ is acquired from the wavelet analysis by computing the phase difference between the velocity time series of the \ti\ and \ha\ lines. The height difference $\Delta z$ is determined from the formation height of the \ti\ and \ha\ lines. To infer the formation height of the \ti\ line, we assumed that the \ti\ line and continuum are formed under local thermodynamic equilibrium (LTE) conditions. Thus, the intensity ratio between the \ti\ line core and the continuum can be substituted for their temperature ratio. The continuum is formed at 0\,km, and the temperature value of 6520\,K is adopted from the FALC model \citep{fon93}. From the intensity ratio between the \ti\ line core and the continuum, we determined the formation height of the \ti\ line as 75\,km, where the temperature is 5580\,K. We assumed the formation height of the \ha\ line core to be 1.5\,Mm \citep{ver81, lee12}. For the mass density and sound speed, we also used values from the FALC model at the specific heights.

\section{Results} \label{results}

We identified wave excitation events related to granules undergoing dynamic changes. These granules collapse, fragment, or submerge. Subsequently, velocity oscillations are detected in the photospheric and chromospheric layers. In Fig.~\ref{fig1}, we show the locations of three of them (hereafter event~1, event~2, and event~3) that were investigated in detail. Event~1 is associated with a collapsing granule, event~2 with a fragmenting one, and event~3 with a submerging one.

\subsection{Dynamically changing granules} \label{subsec:granules}

\begin{figure*}[ht!]
\centering
\includegraphics[width=\hsize]{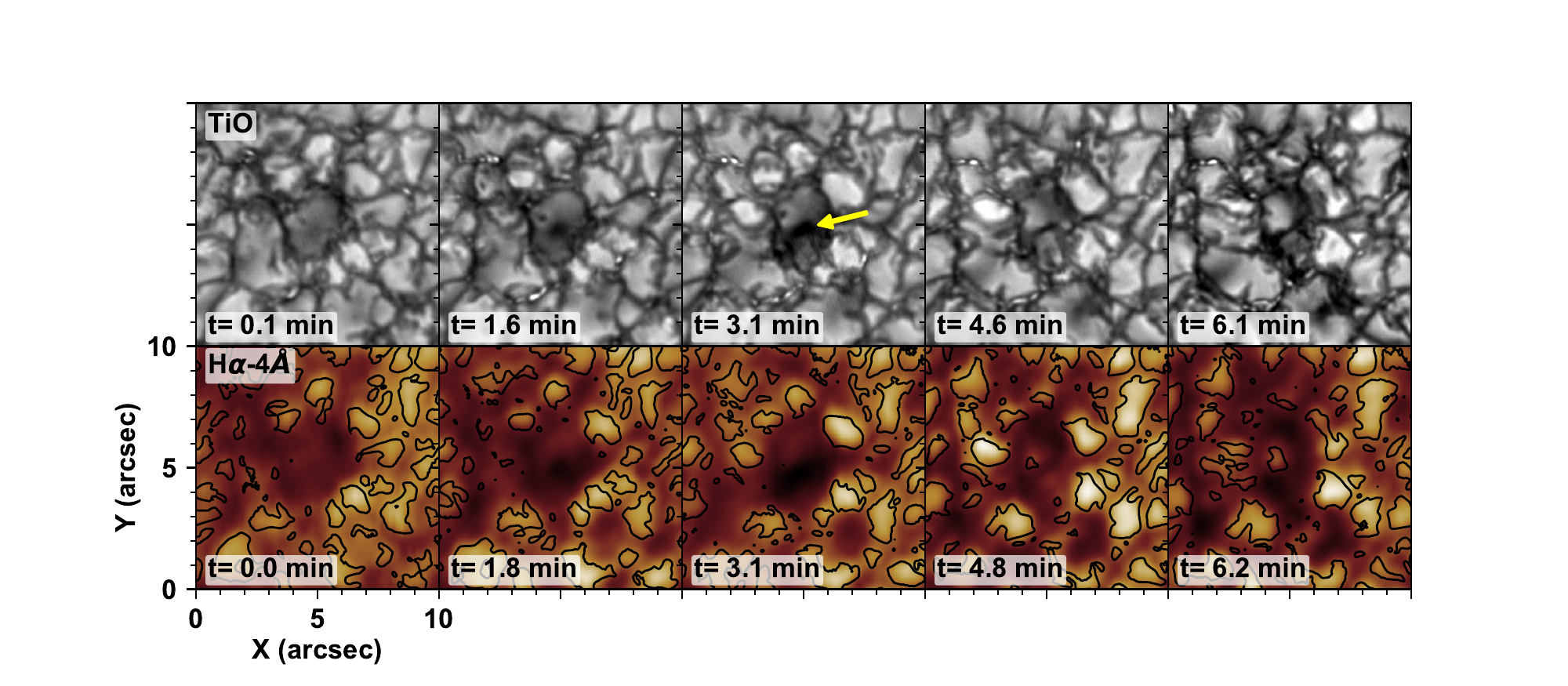}
\caption{Same as Fig.~\ref{th1}, but for event~2. \label{th2}}
\end{figure*}

\begin{figure*}[ht!]
\centering
\includegraphics[width=\hsize]{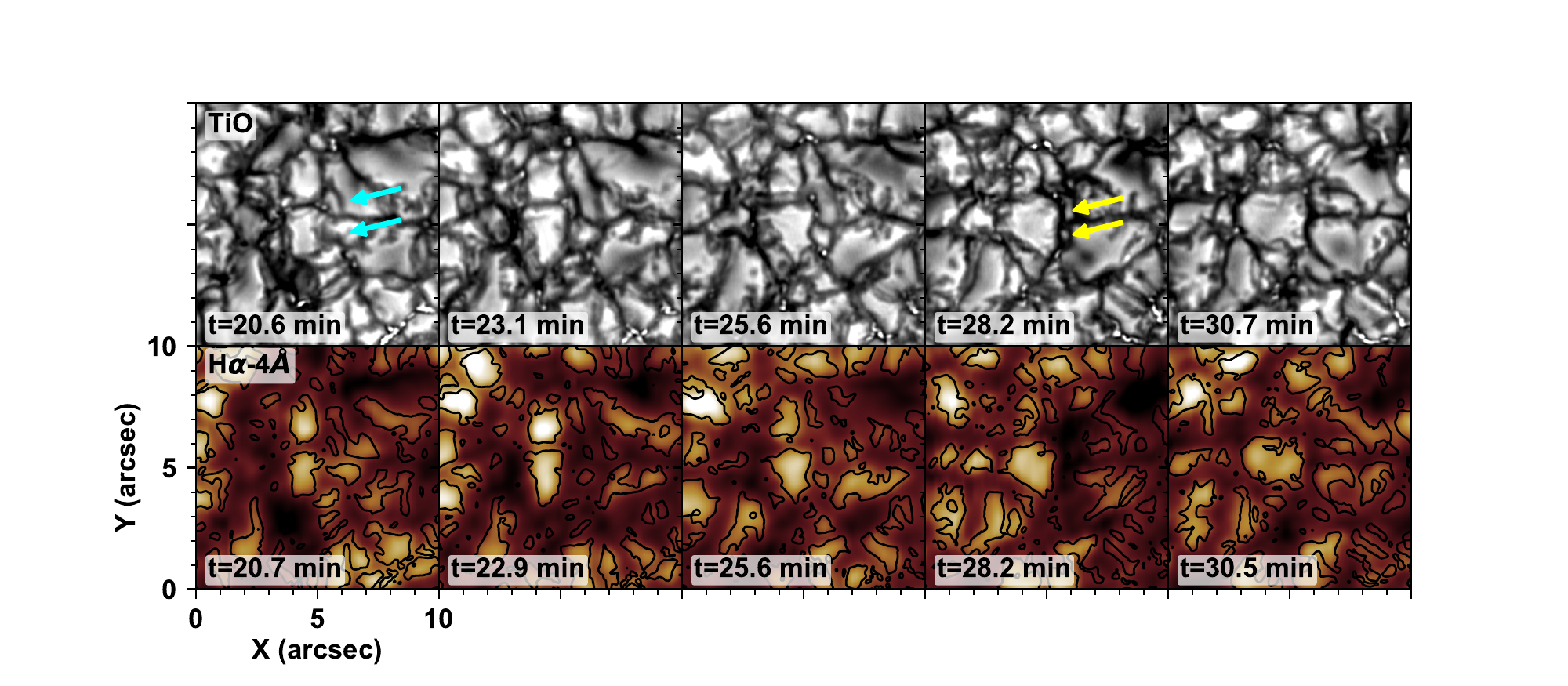}
\caption{Same as Fig.~\ref{th1}, but for event~3. The cyan arrows indicate the initial positions of the submerging granules.\label{th3}}
\end{figure*}

Event~1 represents a case of a granule that became considerably dark with an edge collapsing within 5\,min. Initially ($t=6.1$\,min), in the center of the TiO image, the brightness of the granule was similar to that of the surrounding region, but after 2.5\,min the entire granule began to darken (see Fig.~\ref{th1} and the associated animation (Fig.~\ref{app1})). The granule reached its minimum intensity at $t=13.6$\,min; the intensity was particularly low at the east edge (yellow arrow in Fig.~\ref{th1}). After the collapse, the size of the granule slightly decreased. After $t=13.6$\,min, the granule progressively became brighter and at $t=16.2$\,min returned to its initial brightness. The same granule is also seen in the \ha--4\,\AA\ images, but, due to the limitation of the spatial resolution, the detailed dynamic changes of the granule cannot be followed. The dynamic changes are just seen as a darkening. Likewise, these darkenings are identified in the \ha--4\,\AA\ images of events~2 and 3 (lower panels of Figs.~\ref{th2} and \ref{th3}).

Figure~\ref{th2} shows the time series of event~2, which represents an expanding granule that later splits into two granules. The brightness of the expanding granule is lower than the neighboring granules. At $t=3.1$\,min, the center of the granule became darker (yellow arrow in Fig.~\ref{th2}), while the granule kept expanding with time. At the location where the darkening occurred, the granule split into two individual granules ($t=6.1$\,min).

In the time series of event~3 (Fig.~\ref{th3}), we find two submerging granules (yellow arrows in Fig.~\ref{th3}). Before the submergence ($t=20.6$\,min), these two granules were located (cyan arrows) between the continuously expanding granules (see the time series of TiO images). The two granules shrank with time, whereas surrounding granules kept expanding. Eventually, at $t=30.7$\,min, the small granules submerged and the sites remained as typical intergranular lanes.

\subsection{Wave excitation} \label{subsec:wave}

\begin{figure*}[ht!]
\centering
\includegraphics[scale=0.7]{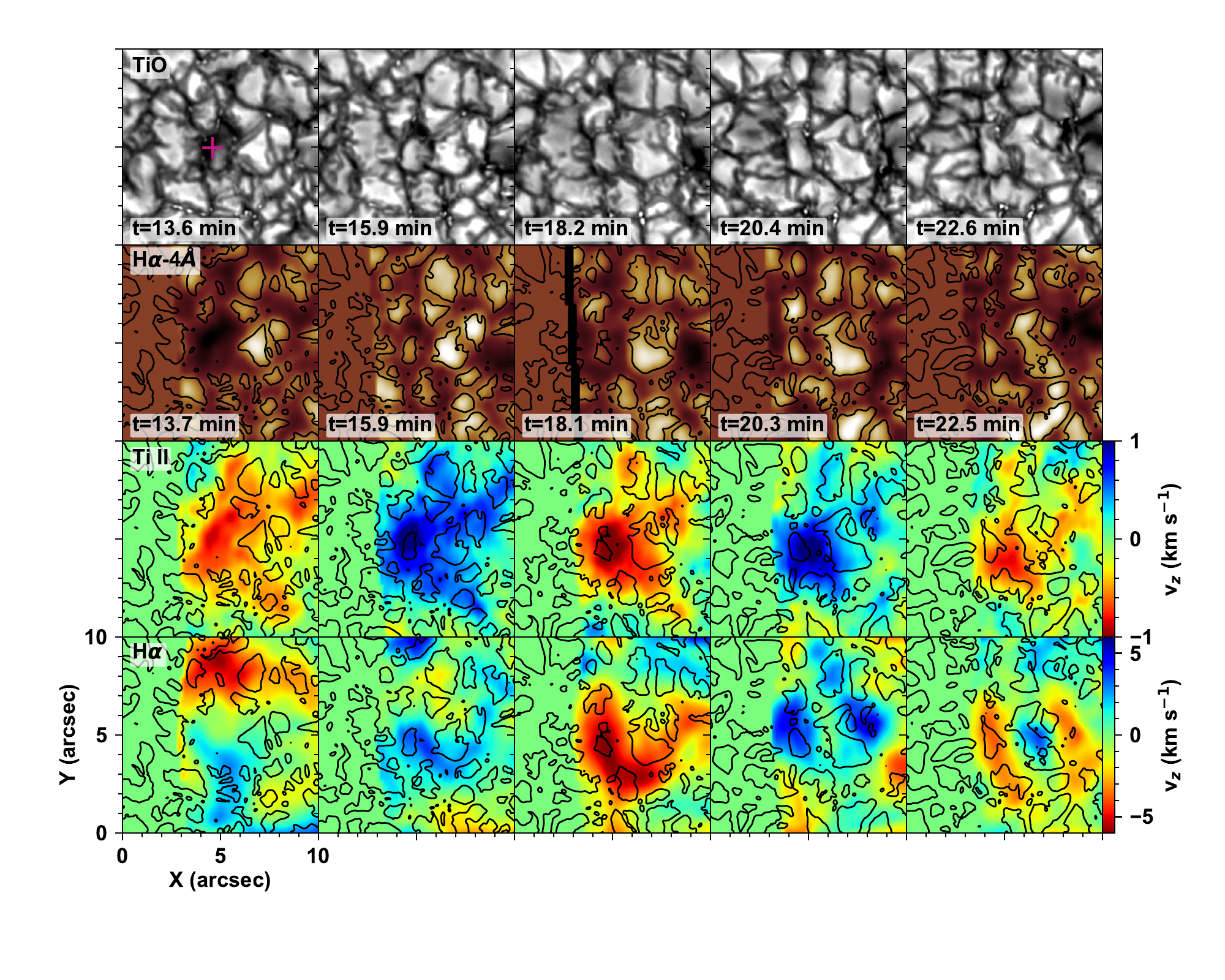}
\caption{Time series of the TiO images (first row), \ha--4\,\AA\ raster images (second row), and vertical velocity maps of the \ti\ (third row) and \ha\ (fourth row). The contours on the \ha--4\,\AA\ raster images and the vertical velocity maps indicate the granules obtained from the TiO images. A purple cross symbol in the first TiO image ($t=13.6$\,min) marks the selected position that was further analyzed and is shown in Figs.~\ref{wavelet} and \ref{flux}. \label{losv1}}
\end{figure*}

Shortly after the dynamic changes of each granule, oscillations are clearly identified in the photosphere and chromosphere. Figure~\ref{losv1} shows the time series of the photospheric images and the Doppler maps of event~1. When the darkening becomes intense ($t=13.6$\,min), the oscillation pattern begins in the \ti\ Doppler maps. A downward velocity patch (red-colored) appears abruptly above the darkening granule and is immediately followed by an upward velocity patch (blue-colored). For a while, downward and upward velocity patches appear in turns. The oscillation pattern lasts approximately 10\,min. The spatial size of the oscillation pattern is approximately 3\farcs5, which is slightly bigger than the size of the granule. A similar oscillation pattern with enhanced amplitude is identified in the \ha\ Doppler maps, but the shape of the oscillation patch is slightly different compared to that in the \ti\ Doppler maps. This discrepancy is not surprising because the photosphere and the chromosphere are not necessarily connected by vertical magnetic fields. The  magnetic configuration is usually complex in quiet regions, and the chromosphere is often permeated by highly inclined magnetic field lines.

\begin{figure*}[ht!]
\centering
\includegraphics[scale=0.7]{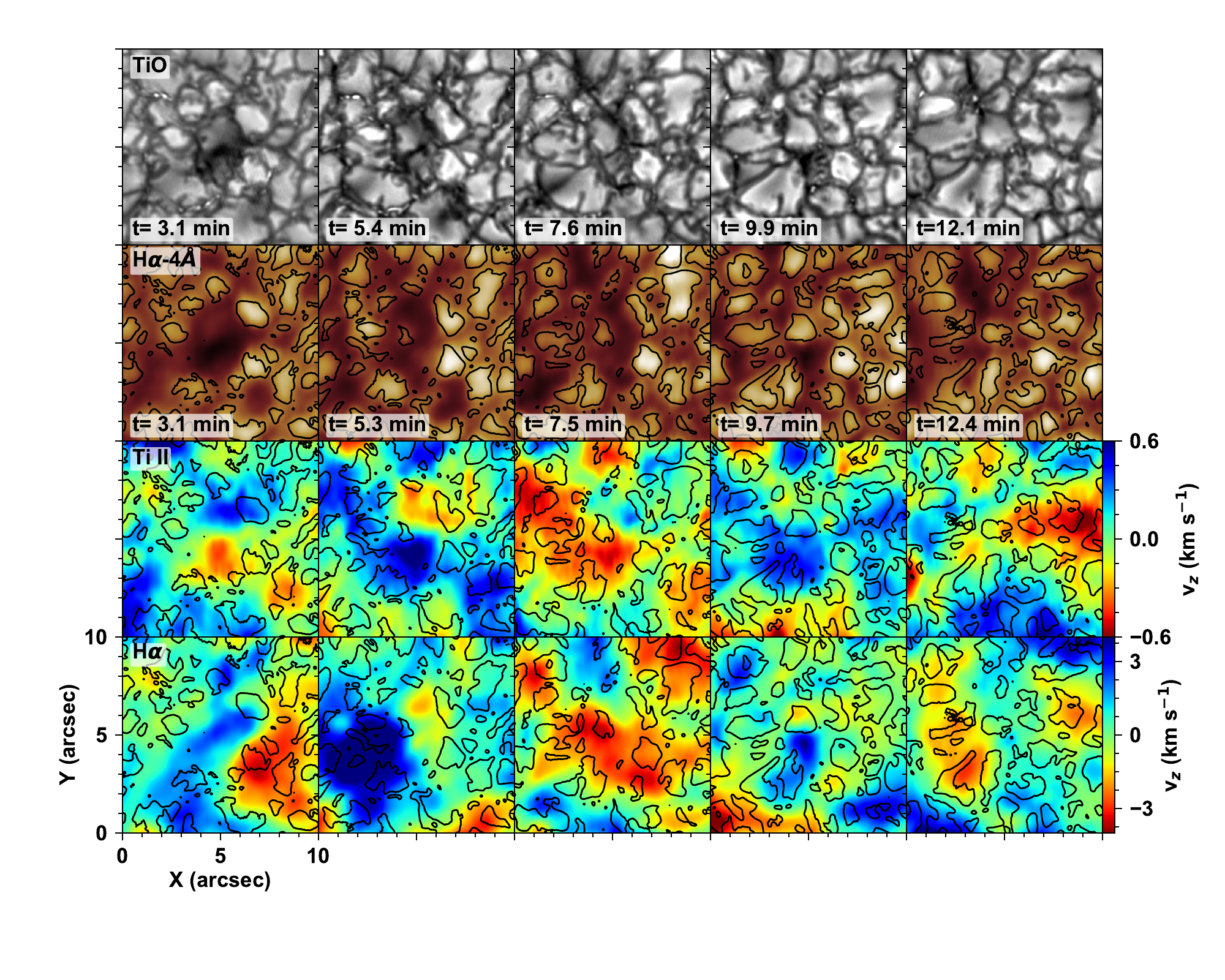}
\caption{Same as Fig.~\ref{losv1}, but for event~2.  \label{losv2}}
\end{figure*}

\begin{figure*}[ht!]
\centering
\includegraphics[scale=0.7]{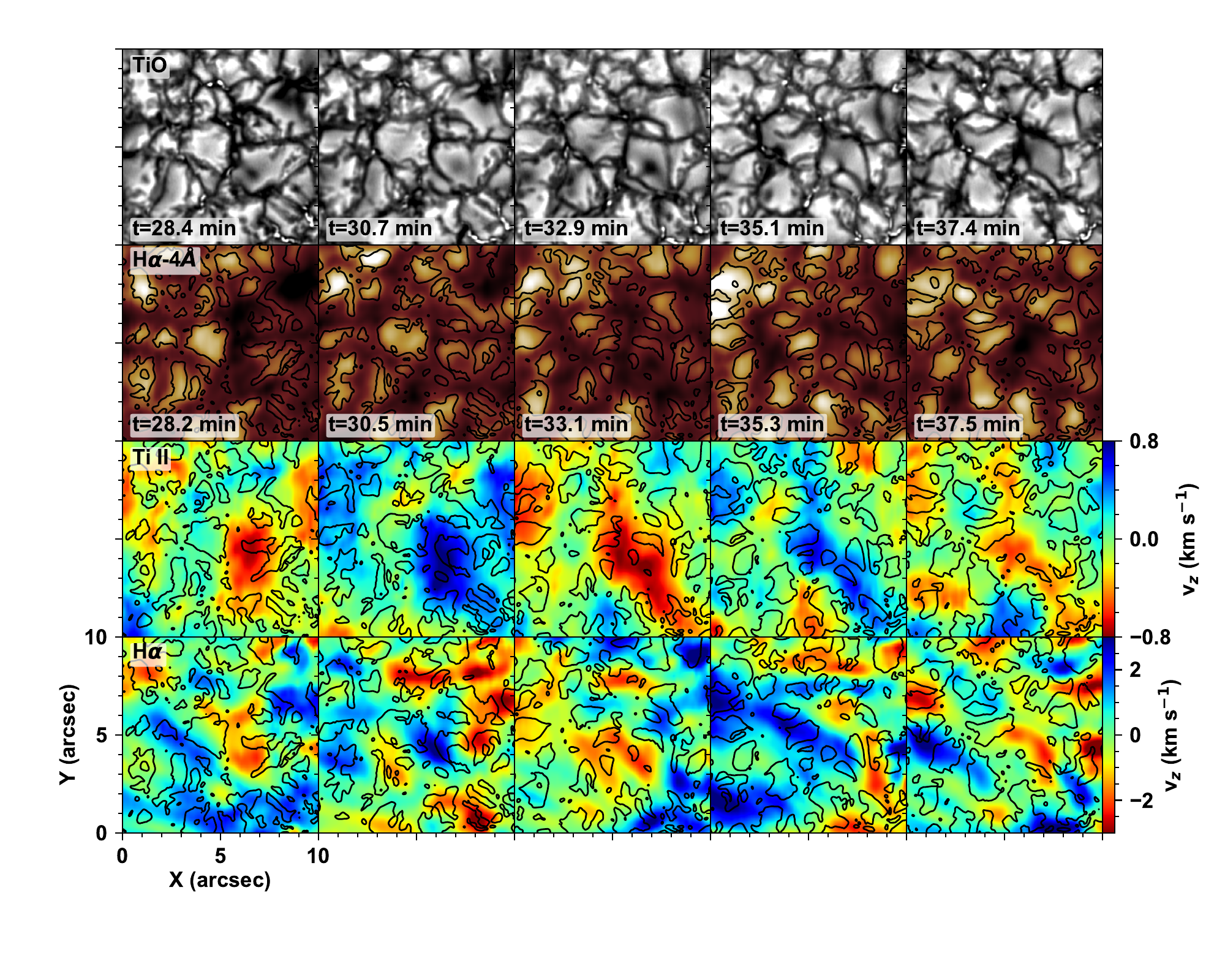}
\caption{Same as Fig.~\ref{losv1}, but for event~3.  \label{losv3}}
\end{figure*}

Likewise, in events~2 and 3, photospheric and chromospheric oscillations are detected above the fragmenting (Fig.~\ref{losv2}) and submerging (Fig.~\ref{losv3}) granules. In the \ti\ Doppler maps, the locations of the velocity patches coincide exactly with those of the changing granules. However, the shapes of the velocity patches are less obvious compared to event~1, probably because of the non-vertical magnetic configuration as described above. Also, the amplitudes of the oscillations are smaller. Therefore, it is difficult to distinguish the oscillation pattern from the surrounding velocity patterns that are not related to the dynamically changing granules. In the \ha\ Doppler maps, the oscillation pattern is hard to notice. Thus, event~1 shows the best isolated oscillation pattern in both the photospheric and the chromospheric Doppler maps. We chose this event for further analysis.

\begin{figure*}[ht!]
\centering
\includegraphics[width=\hsize]{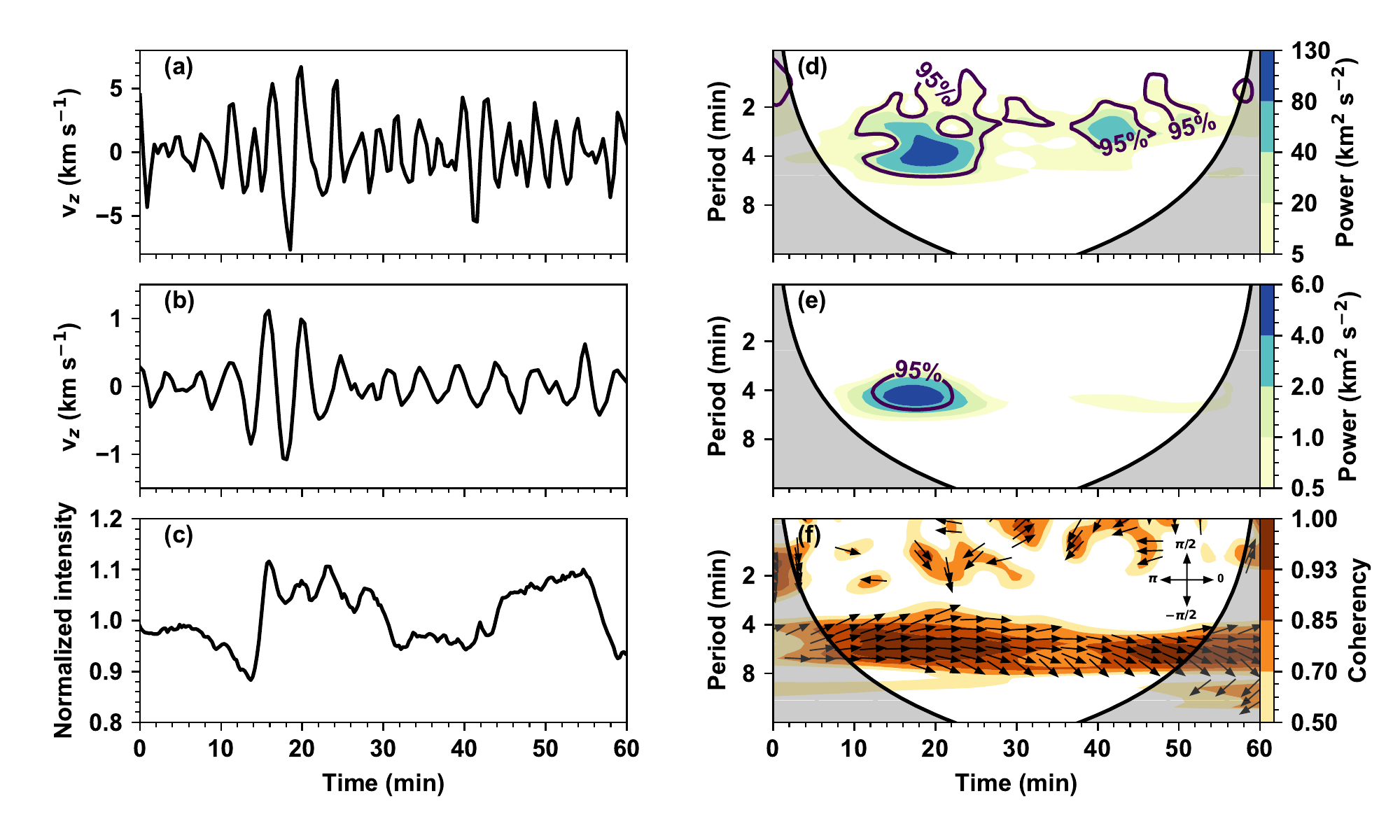}
\caption{Left: temporal variations of the \ha\ vertical velocity (a), \ti\ vertical velocity (b), and TiO intensity (c) at the position marked with a cross symbol in Fig.~\ref{losv1}. Right: wavelet power spectra of \ha\ vertical velocity (d) and \ti\ vertical velocity (e). The purple contours represent the 95\,\%\ local confidence level. The coherency between the two time series of velocity is shown by contours in (f). The phase differences at the points of coherency above 0.7 are denoted by the directions of the arrows. \label{wavelet}}
\end{figure*}

Figure~\ref{wavelet} presents the temporal variations of the intensity and the velocities of event~1. The TiO intensity and the vertical velocities of the \ti\ and \ha\ lines are obtained from a selected position in Fig.~\ref{losv1}. The position is chosen because it is the location at the dark edge of the granule that has the highest photospheric velocity. As mentioned above, the intensity of the granule is close to the spatially averaged value of the intensity at the start of the event. A few minutes later, it starts to gradually decrease in time. The intensity reaches its minimum value at $t=13.6$\,min. For 13.6\,min, the intensity decreases by about 12\,\%. Then, the intensity rapidly increases to a value even higher than the initial value.

The velocity oscillations are identified in the time series of 1--6\,min bandpass-filtered velocities. In the \ti\ line, just before the intensity minimum, the oscillations start with a downward motion. The amplitude of the oscillations is approximately 1\,\kms. The corresponding wavelet power spectrum indicates that the power of the oscillations is mostly concentrated in the range of 3 to 6\,min, and the dominant period of the oscillations is 4.2\,min. In the \ha\ line, the amplitude of the oscillations is about 5\,\kms, and the power is concentrated in a similar range but includes more short-period components than that of the \ti\ line. The dominant period of the \ha\ velocity oscillations is 3.8\,min.

We calculated the wavelet coherency and phase difference of the velocities measured at two different layers. The coherency becomes higher when the granule is sufficiently darkened ($t\sim10$\,min), especially in the 4 to 6\,min period range. In the case of 3 to 4.5\,min period waves, the phase differences are at approximately 0.35--0.5\,rad, which indicates an upward propagation of the waves. In the longer period regime ($>4.5$\,min), the phase differences are close to zero or have negative values. This indicates downward propagating waves. It is considered to be a natural consequence since the longer period waves, which are longer than the cutoff period, cannot go through the chromosphere.

\subsection{Wave propagation and influence on the higher atmosphere} \label{subsec:prop}

\begin{figure}[ht!]
\centering
\includegraphics[width=\hsize]{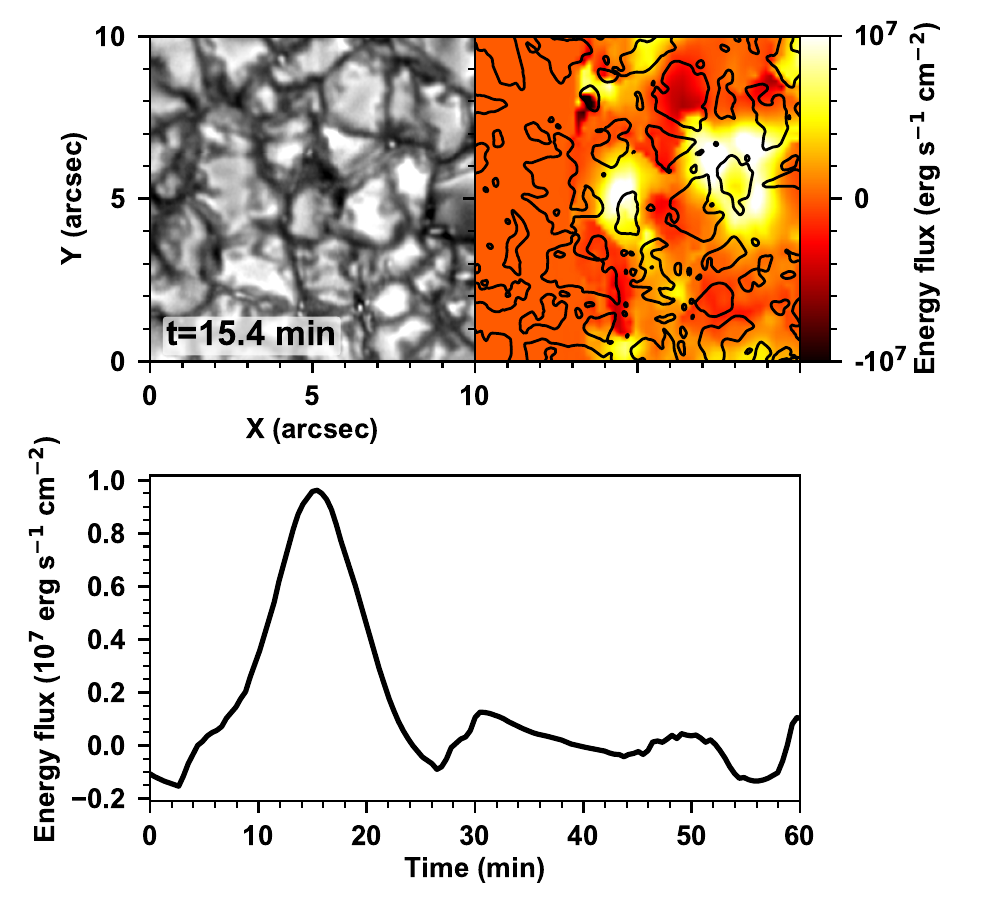}
\caption{TiO image (top left) and photospheric wave energy flux map (top right) at $t=15.4$\,min. The contours on the wave energy flux map indicate the granules obtained from the TiO images. The temporal variation of the wave energy flux (bottom) is acquired at the position marked with a cross symbol in Fig.~\ref{losv1}.  \label{flux}}
\end{figure}

During event~1, upward wave energy flux is enhanced during the photospheric and chromospheric velocity oscillations (see Fig.~\ref{flux}). We measured the wave energy flux integrated over the periods from 2 to 4\,min (3\,min band) in the photospheric layer. The upward wave energy flux is conspicuous above the granule that underwent darkening. The spatial size of the flux patch is larger than that of the granule. Next to the wave energy flux patch that we are interested in, there is another wave energy flux enhancement that is even larger and located in the intergranular lanes. It represents another wave excitation event related to converging flows that cause some granules to submerge. The lower figure shows the temporal variation of the wave energy flux at the selected position in Fig.~\ref{losv1}. The flux is on the order of $10^{6}$\,\flux\  and the peak value is $9.6\times10^{6}$\,\flux\ at $t=15.4$\,min.

At this point, we compare our acoustic events and regular acoustic waves (p-mode oscillations). The oscillations during the time interval $t=30$  to $40$~min (see Fig.~\ref{wavelet}) are identified with the regular acoustic waves. The velocity oscillations during this time interval have smaller amplitudes in both the \ha\ and \ti\ vertical velocity plots and have the phase differences that are close to zero or negative values, meaning that they are standing waves or downward propagating waves. As expected, the wave energy flux is close to zero (see Fig.~\ref{flux}). Therefore, regular acoustic waves cannot transfer energy upward and are distinct from the waves excited by rapidly changing granules.

\begin{figure}[ht!]
\centering
\includegraphics[width=\hsize]{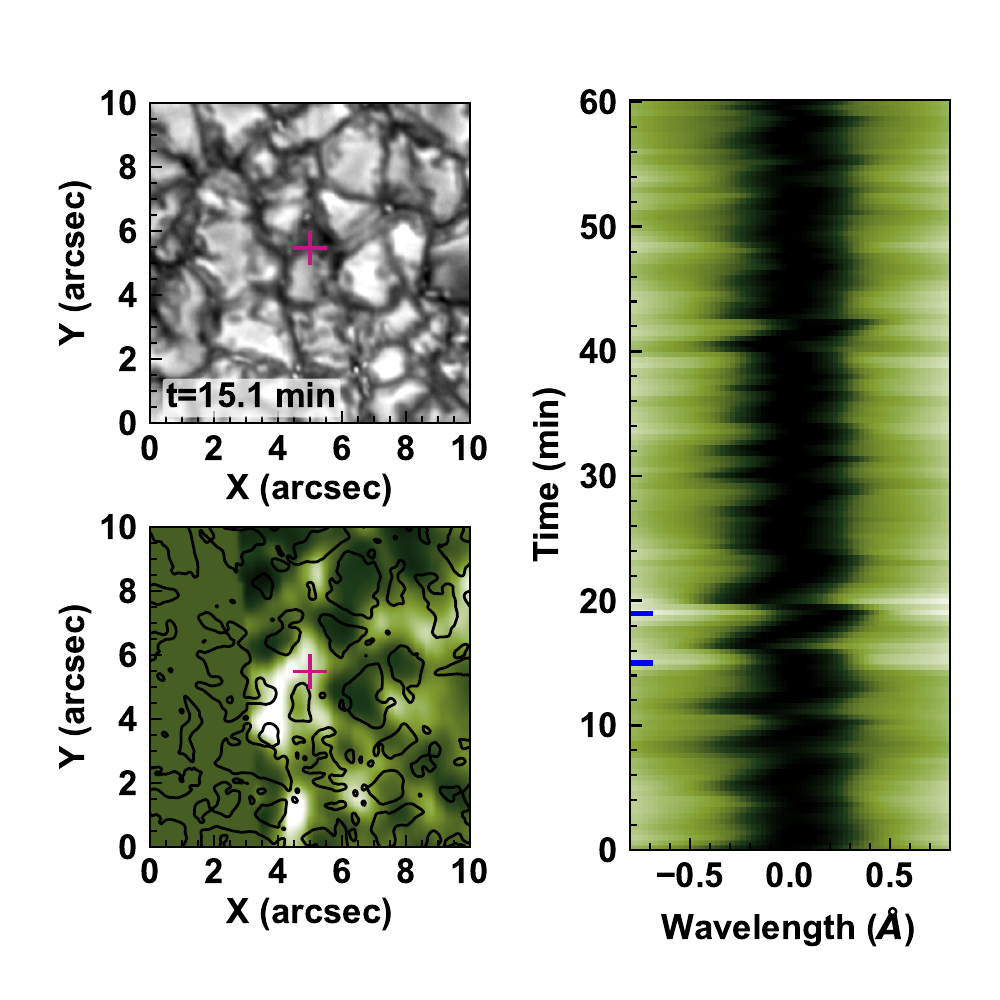}
\caption{TiO image (top left) and the \ca--0.5\,\AA\ raster image (bottom left) at $t=15.1$\,min. The wavelength-time ($\lambda-t$) plot of the \ca~8542\,\AA\ line intensity (right) was taken from the position marked with a purple cross symbol on the \ca--0.5\,\AA\ raster image. In the $\lambda-t$ plot, two brightenings are marked with blue horizontal lines at $t=15$ and $t=19$\,min.  \label{cbp}}
\end{figure}

After the granule darkens ($t=13.6$\,min), repetitive brightenings are identified in the \ca--0.5\,\AA\ raster images (see Fig.~\ref{cbp}). The brightenings occurred at $t=15$ and $19$\,min, and the time interval ($\sim$4\,min) is close to the dominant period of the chromospheric oscillations that were inferred from the \ha\ line. The brightenings are more prominent above the intergranular lanes. The right panel of Fig.~\ref{cbp} shows the wavelength-time ($\lambda-t$) plot of the \ca~8542\,\AA\ line acquired above the brightenings. The collapsing granule event occurred at $t=13.6$\,min, and two brightenings can be identified at $t=15$ and $19$\,min. The duration of each brightening is about 2\,min. The temporal behavior of the brightenings is consistent with previous studies of \ca\ bright grains \citep{cram83, kam06}. In addition, we can clearly see the sawtooth pattern, which is known as the manifestation of shock waves.

\section{Discussion}
\label{discussion}

We have reported the excitation of waves above dynamically changing granules in the internetwork region of the quiet Sun. Three dynamically changing granules that collapse (event~1), fragment (event~2), or submerge (event~3) are studied in detail. Above these granules, spatially localized and transient velocity oscillations are detected in the photospheric and chromospheric layers. In the case of event~1, the dominant periods of the oscillations are 4.2\,min and 3.8\,min in the photosphere and chromosphere, respectively. In addition, upward wave energy flux of up to $9.6\times10^{6}$\,\flux\ is found above the granule. The estimated wave energy flux is more than twice the value of $4\times10^{6}$\,\flux\ required for the chromospheric heating in a quiet Sun \citep{wit77}. The upward propagating waves eventually develop into shocks in the chromosphere. The shocks are identified as repetitive brightenings in the \ca--0.5\,\AA\ images and $\lambda-t$ plot.

Our results suggest that the dynamic changes of granules can trigger transient waves above the granules. This is consistent with the mechanism of acoustic events associated with the darkening of intergranular lanes, which is considered a manifestation of downflows \citep{rim95}. Similarly, the dynamic changes of the granules can lead to downflows. According to simulation studies, the downflows can be induced by localized cooling events \citep{rast99} or by the deficiency of energy that supports the granule \citep{ska00}. The downflowing material rapidly descends below the surface of the photosphere, and when the material reaches a certain depth, it triggers vertical disturbances. \citet{chae15} theoretically demonstrate that an impulsive disturbance in a gravitationally stratified medium produces acoustic waves of frequencies close to the cutoff frequency. The dominant period of the chromospheric oscillations we detected is about 3.8\,min, which is slightly longer than the cutoff period of the internetwork region ($\sim3$\,min). Nevertheless, according to \citet{kay18}, successful propagation of photospheric 5\,min oscillations to the chromosphere happens at some locations in the internetwork region of the quiet Sun, which might be the result of the magnetic field inclination. Our results also indicate that the formation of the \ca\ brightenings are closely related to the 3\,min band photospheric oscillations generated by the darkening granule event.

Is the new type of excitation event we observed physically distinct from the acoustic events reported in earlier studies?  Our new-type events seem to show certain differences from the classic acoustic events. The classic acoustic events are found predominantly in the intergranular lanes \citep{rim95, bel10, mal12}. In our data, we also identified one classic acoustic event located in the intergranular lanes. Its physical properties and the time series are given in Appendix~\ref{cae}. However, our new-type events are not limited to the intergranular lanes. Rather, wave excitations initially occurred in the middle of the granules, where they began to collapse (see Fig.~\ref{losv1}), fragment (see Fig.~\ref{losv2}), or submerge (see Fig.~\ref{losv3}). Furthermore, the spatial extent of the excitation is much larger than that of the classic acoustic events.

Despite these differences, the new-type events do not seem to be physically distinct from the classic acoustic events. Darkenings and intense downflows characterize both types of events. In the case of the classic acoustic events, \citet{rim95} reported that intergranular lanes become darker several minutes before the excitation of waves. They suggested that the darkening reflects localized cooling that occurs in the intergranular lanes and that it is accompanied by downflows. Simulations by \citet{ska00} showed that the vanishing of small granules accompanies downflows and initiates acoustic waves. They noted that these collapsing granules are located at the boundaries of mesogranules, where the upward enthalpy flux is smaller than average. These collapsing granules may be the same type as the rapidly changing granules we observed. We are, therefore, the first to report the observational detection of the rapidly changing granules as the source of wave excitation predicted from the numerical experiment of \citet{ska00}. We note that the same hydrodynamical process can account for both the acoustic events and the rapidly changing granules \citep{ska00}. Both types of events originate from the sudden cooling resulting from the deficit of upward convective energy flux. The cooling accounts for darkening, downflows, and the excitation of waves. In addition, we found that the classic acoustic event and the new-type acoustic events are similar in physical properties, such as the dominant period of the oscillations and the wave energy flux. Since there is no physical difference, our events may be called acoustic events of a new type, and the previously reported events called classic acoustic events. 

There is also the possibility that the two types of acoustic events may not form two distinct groups from the observational point of view. It is likely that some if not all photospheric darkening events or acoustic events reported in earlier studies \citep{rim95, cad03} were in fact acoustic events of a new type. We find more dynamically changing granules in our data, and a follow-up study will explore their statical properties and relationship with the classic acoustic events.

\begin{appendix}
\section{Online material} \label{online}
\begin{figure}[ht!]
\centering
\hspace*{-1cm}
\includegraphics[width=\hsize]{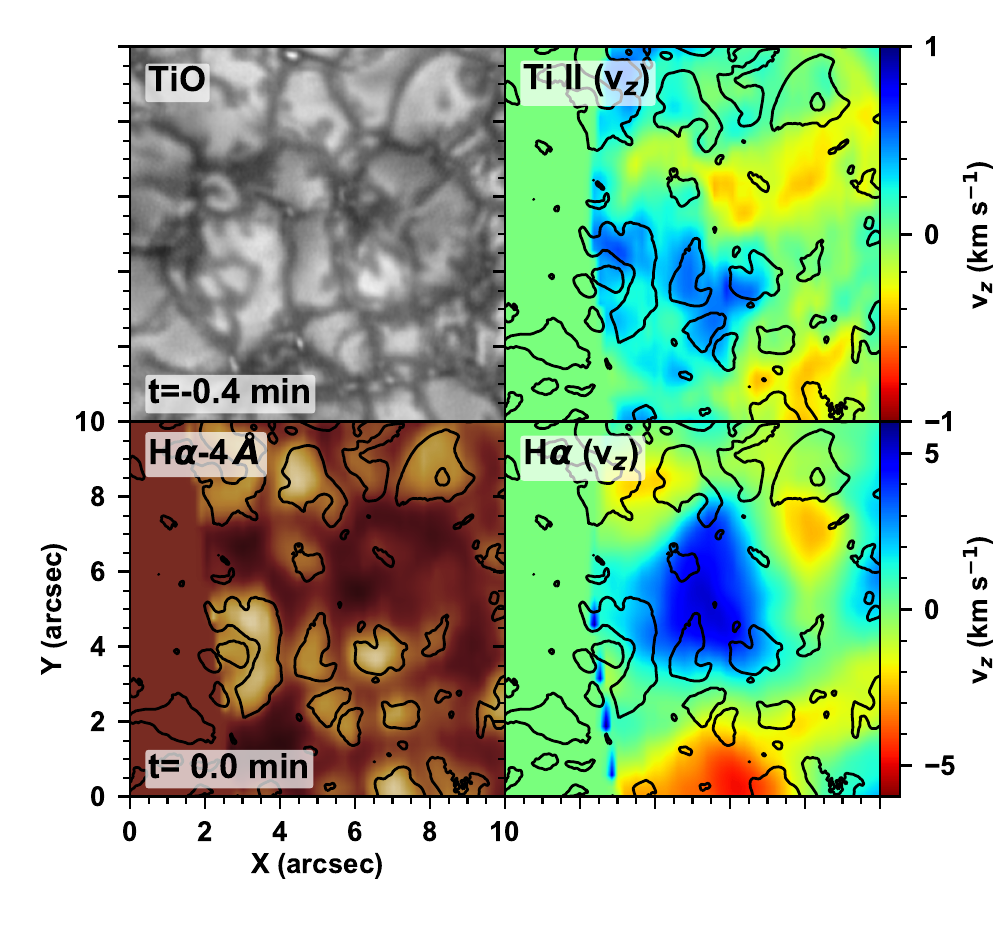}
\caption{Animation sequence of the TiO images (top left), \ha--4\,\AA\ raster images (bottom left), and the \ti\ and \ha\ vertical velocity maps (top and bottom right, respectively). The contours on the \ha--4\,\AA\ raster images and the vertical velocity maps indicate the granules obtained from the TiO images. The animation shows the time series of event~1, including the collapse of the granule and the wave excitation. }
\label{app1}
\end{figure}

\section{Classic acoustic event} \label{cae}
\begin{figure*}[ht!]
\centering
\includegraphics[scale=0.7]{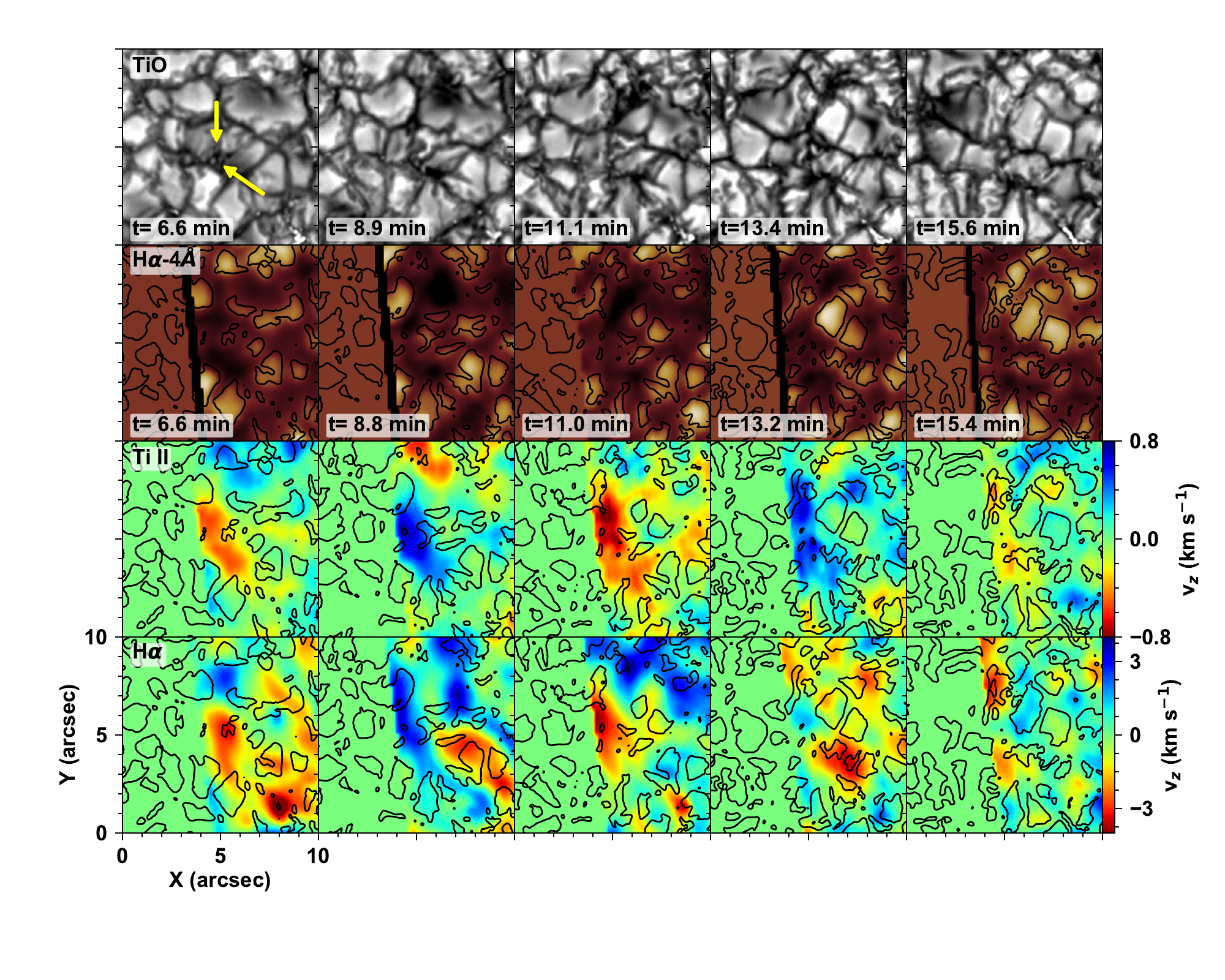}
\caption{Same as Fig.~\ref{losv1}, but for a classic acoustic event. The yellow arrows indicate the position where the classic acoustic event occurred above the intergranular lanes. \label{losvc}}
\end{figure*}
Figure~\ref{losvc} shows the time series of the photospheric and the Doppler maps of a classic acoustic event. At $t=6.6$\,min, a darkening of intergranular lanes is identified in the photospheric images. It is obvious in the \ha--4\,\AA\ raster image. At the same time, an oscillation pattern starts with a downward velocity patch in the \ti\ and \ha\ vertical velocity maps, and the pattern lasts about 10~min. Since the oscillation pattern is confined to the intergranular lanes, the spatial size of the oscillation pattern is smaller than that of the new-type acoustic events reported here. The dominant periods of the oscillation are 4~min at the photospheric level and 3.9~min at the chromospheric level. The wave energy flux above the intergranular lanes is on the order of $10^{6}$\,\flux. The maximum value is $3.6\times10^{6}$\,\flux.

\section{Noise model} \label{model}
\begin{figure*}[ht!]
\centering
\includegraphics[scale=0.7]{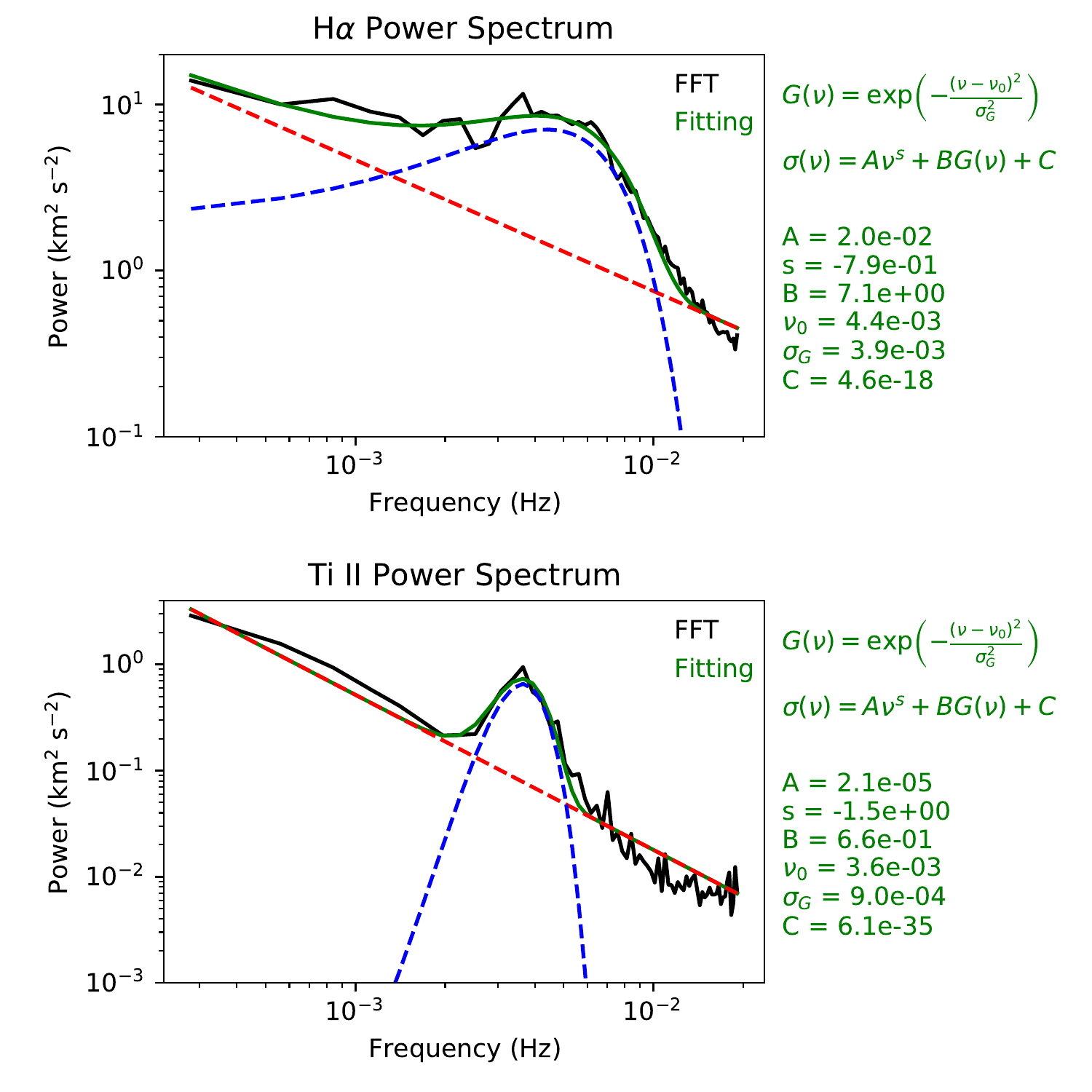}
\caption{Fourier power spectra and noise models for the \ha\ (upper panel) and \ti\ (lower panel) data. In each panel,  the solid black line represents the Fourier power spectra and the solid green line represents their model fitting results. The dashed red and blue  lines indicate the power law and Gaussian term, respectively. \label{noise model}}
\end{figure*}
Figure~\ref{noise model} shows the Fourier power spectra and noise models for the \ha\ and \ti\ data. The models consist of a power law and a Gaussian term. We note that the white noise term is negligible in both cases. Using this information, we calculated the 95\,\% local confidence level for both data (see Fig.~\ref{wavelet}~(d) and (e)). The results clearly show that the identified waves generated by event~1 have wavelet power that exceeds the 95\,\% local confidence level in both lines.

\end{appendix}

\begin{acknowledgements}
We appreciate the referee's constructive comments. This research was supported by the National Research Foundation of Korea (NRF-2020R1A2C2004616), and the Korea Astronomy and Space Science Institute under the R\&D program (Project No. 2020-1-850-07) supervised by the Ministry of Science and ICT, and by the Brain Pool Program of Korea (NRF-2019H1D3A2A0109943). BBSO operation is supported by NJIT and US NSF AGS-1821294 grant. GST operation is partly supported by the Korea Astronomy and Space Science Institute, the Seoul National University, and the Key Laboratory of Solar Activities of Chinese Academy of Sciences (CAS) and the Operation, Maintenance and Upgrading Fund of CAS for Astronomical Telescopes and Facility Instruments.

\end{acknowledgements}

\bibliographystyle{aa}
\bibliography{ms}

\end{document}